\documentclass{aa}  
\usepackage{graphicx}
\usepackage{natbib} 
\bibpunct{(}{)}{;}{a}{}{,}
\usepackage{epsfig}
\usepackage{txfonts}

\begin{document}

\title{A possible activity cycle in \object{Proxima Centauri}}


\author{C. Cincunegui\inst{1,2}$^{,\star}$, R. F. 
D\'{\i}az\inst{1,2}$^{,\star}$, \and P. J. D. 
Mauas\inst{1,3}$^{,}$\thanks{Visiting Astronomers, Complejo Astron\'omico El 
Leoncito, operated under agreement between the Consejo Nacional de 
Investigaciones Cient\'{\i}ficas y T\'ecnicas de la Rep\'ublica Argentina and 
the National Universities of La Plata, C\'ordoba and San Juan.}}

\offprints{C. Cincunegui, \email{carolina@iafe.uba.ar}}

\institute{Instituto de Astronom\'{\i}a y F\'{\i}sica del Espacio, 
C.C. 67 Suc. 28 (1428) Buenos Aires, Argentina \and Fellow of the CONICET \and 
Member of the Carrera del Investigador Cient\'{\i}fico, CONICET}

\titlerunning{A possible activity cycle in Prox Cen}

\authorrunning{C. Cincunegui, R. F. D\'iaz, \and P. J. D. Mauas}

\date{Received <date>; accepted <date>}

\newcommand{\hal}{\ensuremath{$H$\alpha}}
\newcommand{\ca}{\ensuremath{$Ca$ $~II$ }}
\newcommand{\hel}{\ensuremath{$D$_3}}
\def\teff{\hbox{$T_{\mathrm{eff}}$}}
\def\deg{\hbox{$^\circ$}}
\def\sun{\ensuremath{\odot}}
\newcommand{\prot}{\ensuremath{P_\mathrm{{rot}}}}

\abstract
{Several late-type stars present activity cycles resembling the Solar one. 
This fact has been observed mostly in stars ranging from F to K, 
i.e., in stars with a radiative core and an outer convective layer.} 
%
{This work aims at studying whether an activity cycle can be detected in the dM5.5e star
Proxima Centauri, which is supposed to be completely convective.}
%
{We present periodical medium-resolution echelle observations 
covering the complete visual range, which were taken at the CASLEO Argentinean Observatory. 
These observations are distributed over 7 years. We discarded the spectra that 
present flare activity, and analyze the remaining activity levels using four 
different statistical techniques to look for a period of activity.} 
%
{We find strong evidence of a cyclic activity, with a period of $\sim$442 
days. We also estimate that the \ca\ $S$ index varies around 130\% due to 
activity variations outside of flares.}
%
{}


\maketitle

%

\section{Introduction}

It is generally accepted that magnetic activity in late-type stars, and in particular 
activity cycles like the one observed in the Sun, are the product of the $\alpha \Omega$ 
dynamo, which results from the action of differential rotation at the 
tachocline (the interface between the convective envelope and the radiative 
core). Therefore, the presence and characteristics of activity cycles 
are closely related to the existence and depth of an outer convection zone. 
Since this depth depends on spectral type --~from F stars that have shallow 
convection zones to middle M stars that are totally convective~--, it is of 
special interest to study these cycles in stars of different spectral types, and 
in particular in middle-M stars, to determine whether there is an onset of 
cyclic activity. 

To date, activity cycles have been detected in several late-type stars \citep[see, 
for example,][]{1985ApJ...294..310B}, mainly measuring variations in the well 
known $S$ index, essentially the ratio of the flux in the core of the Ca II H 
and K lines to the continuum nearby. This index has been defined at the Mount 
Wilson Observatory, where an extensive database of activity has been built over 
the last three decades. However, these observations are mainly concentrated on 
stars ranging from F to K \citep[see, for example,][]{1995ApJ...438..269B}, due 
to the longexposure times needed to observe the Ca II lines in the red and faint 
M stars. As a contribution to the subject, we explore the existence of an 
activity cycle in the dM5.5e star Proxima Centauri, using observations 
we made at the CASLEO Observatory, in Argentina and that span 7 years. 

The triple star system $\alpha$~Cen is located at 1.3~pc from the Sun. 
One of its members is Proxima Centauri ($\alpha$~Cen~C; GJ551), 
the closest star to Earth, which has a visual magnitude of 11.01. 
Prox Cen has an effective temperature of 3042~K \citep{2003A&A...397L...5S}, 
a radius of $\sim 1/7\, R_\sun$ \citep{1980A&A....82...53P}, and a mass of $\sim 
0.2\, M_\sun$ \citep{1976asqu.book.....A}. 
Its rotation period has been estimated several times. From photometric observations, 
\citet{1998AJ....116..429B} found \prot=83.5 days, while 
\citet{1987MNRAS.224P...1D} measured \prot=$51\pm12$~days from chromospheric 
activity and \citet{1996AAS...188.7105G}, using IUE observations of \emph{h} and 
\emph{k} Mg~II lines, found \prot=$31.5\pm1.5$~days.
These rotation periods are longer than in others M stars
\citep[see, for example,][]{2003ApJ...586..464B}.
This longer period is consistent with the older age of Prox Cen. If 
Proxima is coeval with its two companions, its age should be 4-4.5~10$^9$~years 
\citep{1986ApJ...300..773D}. 

\begin{table*}[t!]
\caption{Log of the observations. Column 1: a label used in the 
figures; Col.~2: xJD=JD$-$2\,451\,000, where JD is the 
Julian date; Col.~3: exposure time in minutes; Col.~4: spectrum clean of 
cosmic rays (`y') or not (`n') (whether it can be used in the \ca\ and \hal\ 
windows); spectra indicated with `--' do not include the corresponding 
region, those with `$\star$' are the ones not used in 
Sect.~\ref{sec:halfa}, due to the presence of flares.} 
\label{tab:obs}
\begin{center}
 \begin{tabular}{|lccc|c|lccc|c|lccc|} \cline{1-4} \cline{6-9} \cline{11-14}                                    
 label &  xJD  &  $t$  & Ca/\hal & & label &  xJD  &  $t$  & Ca/\hal & & label &  xJD  &  $t$  & Ca/\hal \\ \cline{1-4} \cline{6-9} \cline{11-14}  
 0399a & 241.85    & 25 & --/y & & 0303a $\star$ & 1\,715.72 & 45 & y/y & & 0904a & 2\,274.48 & 30 & y/y \\ 
 0399b & 241.87    & 25 & --/y & & 0303b & 1\,715.75 & 45 & y/y & & 0904b & 2\,274.50 & 30 & y/y \\ 
 0399c & 241.89    & 15 & --/n & & 0303c & 1\,716.60 & 60 & y/y & & 0305a & 2\,448.78 & 45 & y/y \\ 
 0399d & 242.67    & 30 & y/-- & & 0303d & 1\,716.65 & 60 & y/y & & 0305b $\star$ & 2\,448.81 & 45 & y/y \\ 
 0399e & 242.70    & 30 & y/-- & & 0303e & 1\,716.69 & 60 & y/y & & 0605a & 2\,523.55 & 60 & y/y \\ 
 0300a $\star$ & 626.82    & 30 & y/y  & & 0303f & 1\,716.74 & 60 & y/y & & 0605b & 2\,523.60 & 60 & y/y \\ 
 0300b $\star$ & 626.84    & 30 & y/y  & & 0303g & 1\,716.78 & 60 & y/y & & 0605c & 2\,523.64 & 60 & y/y \\ 
 0300c $\star$ & 626.86    & 30 & y/y  & & 0303h & 1\,716.82 & 60 & y/y & & 0605d & 2\,523.69 & 60 & y/y \\ 
 0300d $\star$ & 626.88    & 30 & y/y  & & 0303i & 1\,716.87 & 60 & y/y & & 0605e & 2\,523.74 & 60 & y/y \\ 
 0300e & 627.82    & 45 & y/y  & & 0603a & 1\,804.59 & 45 & n/n & & 0605f & 2\,523.78 & 60 & y/y \\ 
 0300f & 627.86    & 45 & y/y  & & 0603b & 1\,804.63 & 45 & n/n & & 0605g & 2\,543.60 & 90 & n/y \\ 
 0301a & 972.78    & 60 & y/y  & & 0903a & 1\,895.51 & 33 & y/y & & 0605h & 2\,543.66 & 90 & n/y \\ 
 0301b & 972.83    & 60 & y/y  & & 0903b & 1\,895.53 & 33 & y/y & & 0805a & 2\,600.51 & 80 & y/y \\ 
 0301c & 974.70    & 90 & y/y  & & 0604a & 2\,160.59 & 45 & n/y & & 0805b & 2\,600.57 & 80 & y/y \\ 
 0301d & 974.77    & 90 & y/y  & & 0604b & 2\,160.63 & 45 & n/y & & 0206a $\star$ & 2\,780.80 & 80 & y/y \\ 
 0701a & 1\,096.46 & 90 & y/y  & & 0604c & 2\,161.47 & 60 & n/n & & 0206b & 2\,780.86 & 80 & y/y   \\ \cline{11-14} 
 0701b & 1\,096.53 & 90 & y/y  & & 0604d & 2\,161.51 & 60 & y/y & & Total &           &    & 49/53 \\ \cline{11-14} 
 0302  & 1\,364.69 & 45 & y/y  & & 0604e & 2\,161.55 & 60 & y/y \\             
 0602a & 1\,451.66 & 45 & y/y  & & 0604f & 2\,161.61 & 60 & y/y \\             
 0602b & 1\,451.69 & 45 & y/y  & & 0604g & 2\,161.65 & 60 & y/y \\             
 0802a & 1\,519.50 & 45 & y/y  & & 0604h & 2\,161.69 & 60 & y/y \\             
 0802b $\star$ & 1\,519.53 & 45 & y/y  & & 0604i & 2\,161.73 & 60 & n/n \\ \cline{1-4} \cline{6-9}
 \end{tabular}                                    

\end{center}
\end{table*}

Prox Cen is also a very active star: it is the first star where evidence of a 
stellar corona outside flares was found \citep{1980ApJ...236L..33H}.
Several flares have been observed in this star, not only in X rays 
\citep{1988ApJ...328..256R, 2004A&A...416..733R, 2002ApJ...580L..73G, 
2004A&A...416..713G}, but also in UV \citep{1977ApJ...213L.119H}, and 
simultaneously in both  \citep{1989A&A...223..241B,1990A&A...232..387H}. Some 
photometric and spectroscopic flares have been detected 
\citep{1981MNRAS.195.1029W,1994IBVS.4048....1P,1998ASPC..154.1212B} and 
\citet{2003PASA...20..257S} detected flares in radio waves. 

\citet{1990A&A...232..387H} suggested that Proxima presents an activity cycle. 
Using UV observations, \citet{1996AAS...188.7105G} 
observed that short-term activity variations have a minimum around 1995, and 
deduced from this fact that the activity cycle had a minimum at that year, 
although during that time four flares were detected. 
\citet{1998AJ....116..429B} deduced the existence of an activity cycle with a 
period of about 1100~days from photometric observations . 

This paper is organized as follows: In Sect.~\ref{sec:observaciones} we 
present our observations; in Sect.~\ref{sec:actividad} we discuss possible 
activity indicators. We analyze whether our activity measurements are
compatible with cyclic variations in Sect.~\ref{sec:halfa}.
Finally, we summarize and discuss our results in Sect.~\ref{sec:conclusiones}.


\begin{figure*}[t!]
\begin{center}
\begin{tabular}{ccc}
\epsfig{figure=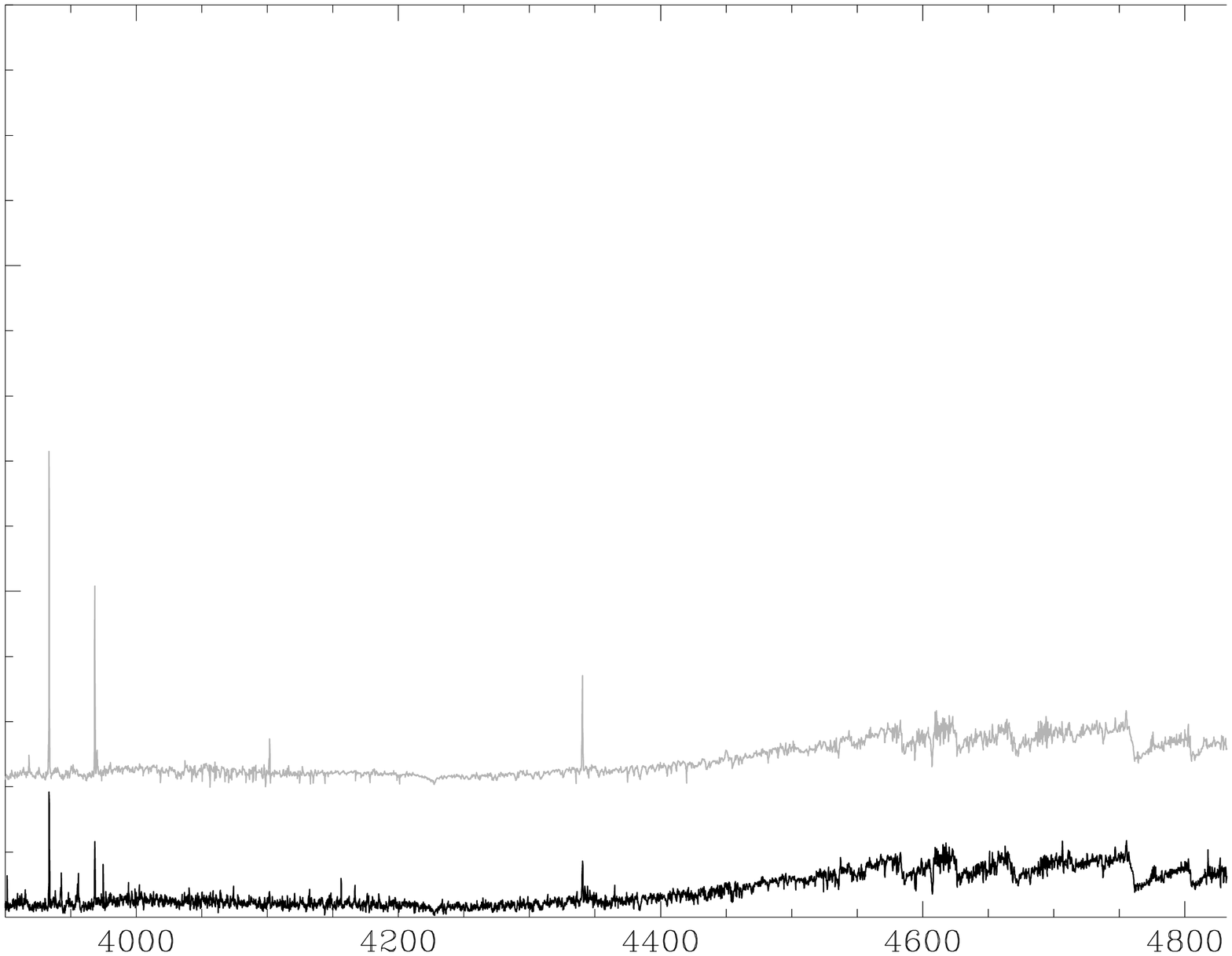,width=49mm,angle=-90}
\hspace{-2mm} 
\epsfig{figure=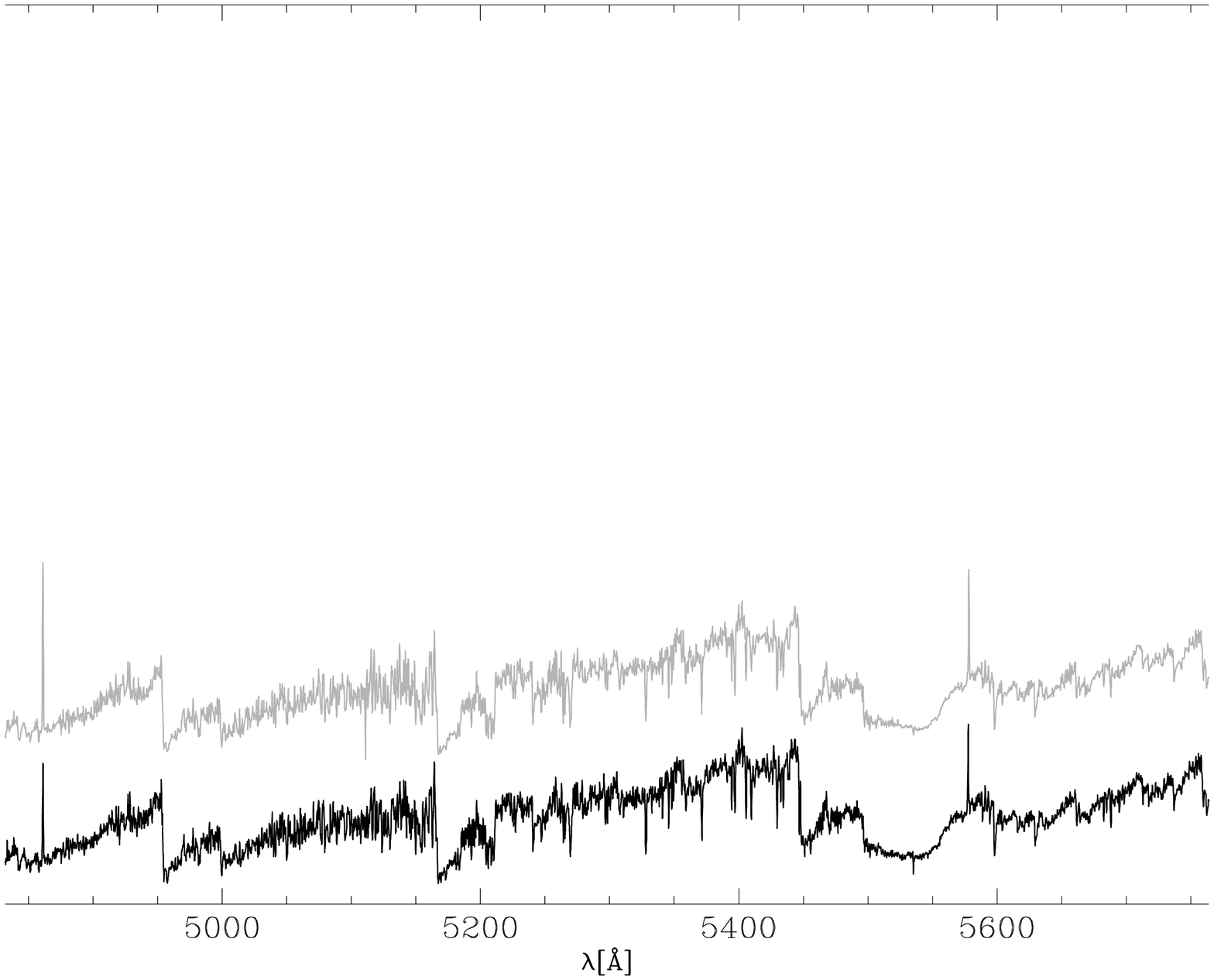,width=49mm,angle=-90}
\hspace{-2.5mm} 
\epsfig{figure=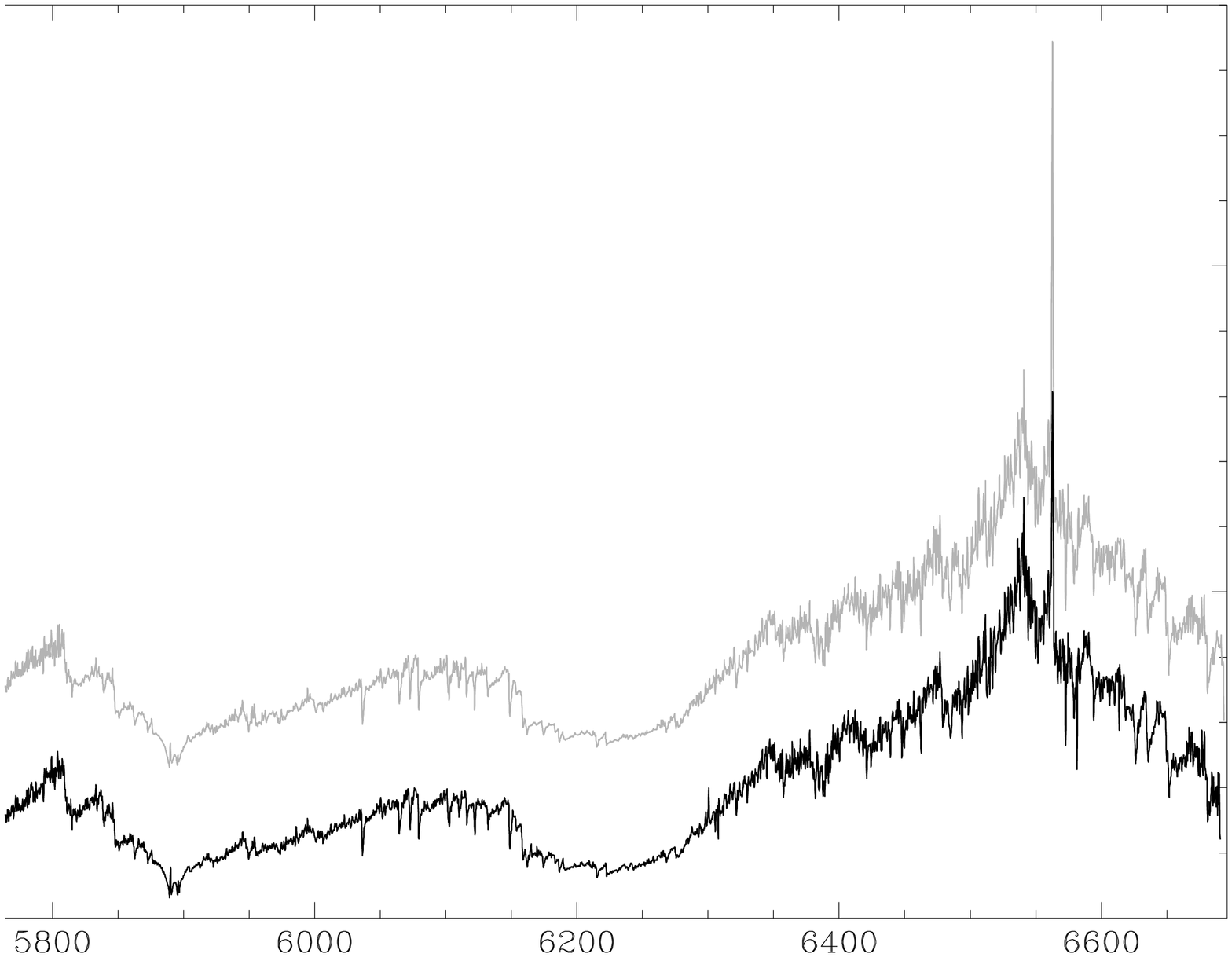,width=49mm,angle=-90}
\end{tabular}
\end{center}
\caption{In this figure we show two of the spectra with the best signal-to-noise 
relations for this star. Each of them is the combination of seven spectra with 
an exposure time of one hour. The upper spectrum was obtained in March 2003 
(xJD 1716) and the one in the bottom in June 2004 (xJD 2161). 
They are arbitrarly displaced for clarity. It can be seen that the spectrum at 
the top shows a larger activity level.} 
\label{fig:espectros} 
\end{figure*} 

\section{The observations}\label{sec:observaciones}

Our observations were made at the 2.15~m telescope of the Complejo 
Astron\'omico El Leoncito (CASLEO), which is located at 2552~m above sea 
level, in the Argentinean Andes. We used a REOSC spectrograph to obtain 
medium-resolution echelle spectra covering the wavelength range  
3900-7000~\AA\ at a spectral resolution of 
R=$\lambda/\Delta\lambda\approx$26400. 
We process the observations according to the method outlined in 
\citet{2004A&A...414..699C}, obtaining flux-calibrated spectra. In 
Table~\ref{tab:obs} we show the observation logs of Prox Cen. The first 
column includes a label used in the figures, built from the month and year of 
the observation, and a letter to distinguish between different spectra from 
the same observation run; the second column lists xJD=JD$-$2\,451\,000, 
where JD is the Julian date at the begining of 
the observation and the third column the exposure time (in minutes) of each 
observation. We inspected the regions of interest of each spectrum to determine 
if they were clean of cosmic rays: in the last column we indicated with `y' or 
`n' whether the spectra is suitable or not to be used in the windows 
corresponding to \ca\ and \hal , as explained in the next section. With `--' we 
indicate some spectra that do not include the region of interest due to a 
different instrumental configuration.

There is a total of 60 individual observations, which have been carried out on 
24 nights distributed over 7 years. In particular, on two of the nights we 
observed Proxima for seven hours, obtaining spectra with very high 
signal-to-noise ratios for this faint star, which are shown in 
Fig.~\ref{fig:espectros}. The upper spectrum corresponds to the 
observations on the night of March 17th, 2003 (spectra 0303c to 0303i), and the 
lower one was obtained the night of June 6th, 2004 (spectra 0604c to 0604i). 
Although no flares occurred on either night, as we will show, the overall 
activity level of the first one is much higher than that of the second one, 
as can be noted by the intensity of several chromospheric lines, in particular 
the \ca\ and the Balmer lines, seen in emission in this dMe star. 

\begin{figure}[b!]
\begin{center}
\epsfig{figure=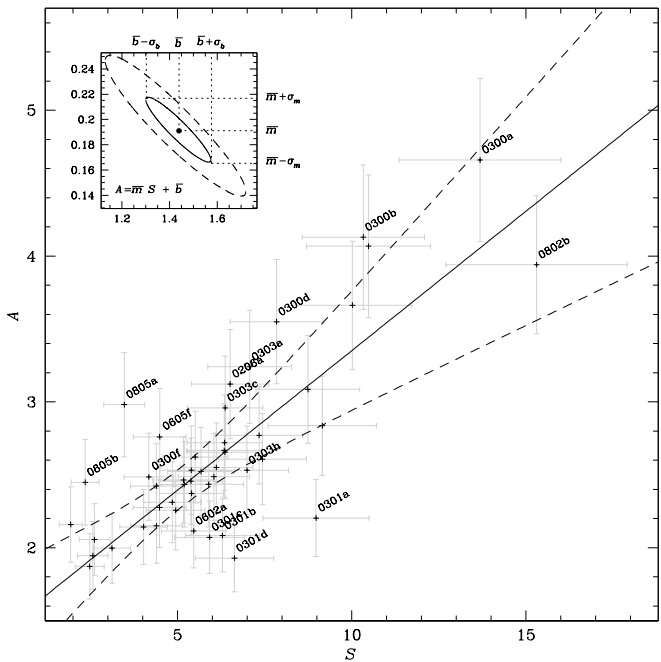,width=\columnwidth}
\end{center}
\caption{Comparison of the \ca\ index, $S$, and the \hal\ index, $A$, for the 
47 spectra in which both spectral features have been measured. The errors are 
assumed to be 12\% in $A$ and 17\% in $S$. The least-squares fit has a slope of 
$0.19\pm0.03$ and a correlation coefficient of -0.94. The fit significance is 
60\%. The dashed lines represent the points that deviate $\pm 3\sigma$ from 
the fit. In the small figure the confidence regions of 39.3\% (full line) and 
90\% (dashed line) are drawn.} 
\label{fig:AvsS} 
\end{figure}
\setcounter{figure}{2}


\section{Activity measurements}\label{sec:actividad} 

The widely-used $S$ index from the Mount Wilson Observatory is the most common 
and best known chromospheric activity indicator for late-type stars. It is 
obtained from measurements taken with a four-channel spectrometer: two of 
the channels are centered in the cores of the H and K \ca\ lines, with a 
triangular profile of 1.09~\AA\ FWHM, and the other two channels acquire the 
flux in the continuum nearby, in two passbands 20~\AA\ wide. The $S$ index is 
built as the ratio between the chromospheric H and K fluxes and the photospheric 
continuum fluxes \citep{1998csss...10..153B}.

To reproduce Mount Wilson's $S$ index we integrated the fluxes in our spectra 
around the K and H \ca\ lines, weighting them with a triangular profile 
resembling the instrumental response of the Mount Wilson spectrometer 
\citep{1978PASP...90..267V}. We normalized these fluxes by the average flux in 
the same two 20~\AA\ continuum passbands used at Mount Wilson, centered at 3901 
and 4001~\AA . In a previous work \citep{2002scsw.conf...91C}, we checked the 
accuracy of our measurements observing a set of 18 non-variable stars given by 
\citet{1996AJ....111..439H}, and using them to intercalibrate our measurements 
with Mount Wilson's $S$. We have found a very good correlation between both 
sets. 
However, Proxima Centauri is a very red and faint star \citep[$B-V$=1.807 
and V=11.01,][]{1997A&A...323L..49P}. Therefore, the \ca\ lines are very hard to 
observe with a good signal-to-noise ratio, requiring very long integration times 
with a 2m class telescope. 

To overcome this problem, it is worth noting that in Proxima Centauri all 
the other Balmer lines, and in particular \hal , are observed in emission, as 
is the case for the most active late-type stars, the dKe and dMe stars. 
Therefore, we constructed an index $A$ from the H$_\alpha$ profile, defined as 
the ratio between the average flux in a 1.5~\AA\ square passband centered in the 
line and the average flux in a 20~\AA\ nearby continuum window, centered at 
6605~\AA , in a similar way to the definition used for $S$. 

To check whether $A$ can be used as an activity index for this red 
star, we explored the relation between $A$ and our $S$ index, which is shown in 
Fig.~\ref{fig:AvsS}. We estimated the errors in the flux-calibration method 
employed at 10\% \citep{2004A&A...414..699C}. Therefore, 10\% should be 
considered as a lower limit for the error in each index. 
Using a non-linear $\chi^2$ minimization to fit a straight line 
considering the errors in both coordinates, and varying the percentual 
errors for each index until finding a reasonable significance of 60\%, we found 
a very good correlation between both indexes, with a correlation coefficient of 
0.94 and a slope of 0.19 for errors of 17\% in $S$ and 12\% in $A$. The 
difference between the errors is natural since the \ca\ lines have a much lower 
signal-to-noise ratio. In the insert to Fig.~\ref{fig:AvsS}, we show the 
confidence regions for the parameters of the fit. 
In spite of the much lower sensitivity of H$_\alpha$, as indicated from the 
small slope, we chose the $A$ index to study activity in this star because of 
the much better signal-to-noise ratio. 

\begin{figure*}[h!]
\begin{tabular}{c}
\psfig{figure=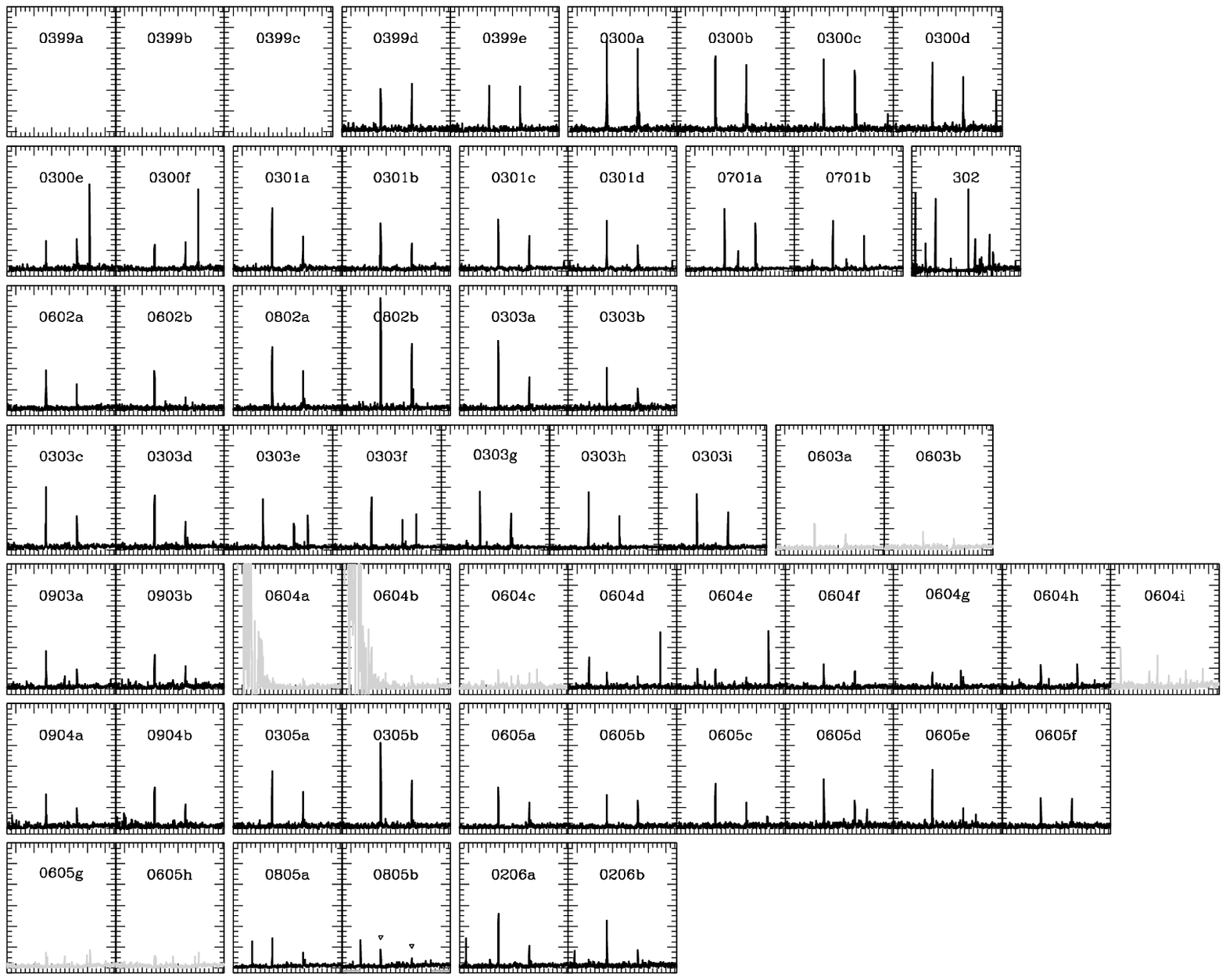,width=0.82\textwidth,angle=0} \\
\psfig{figure=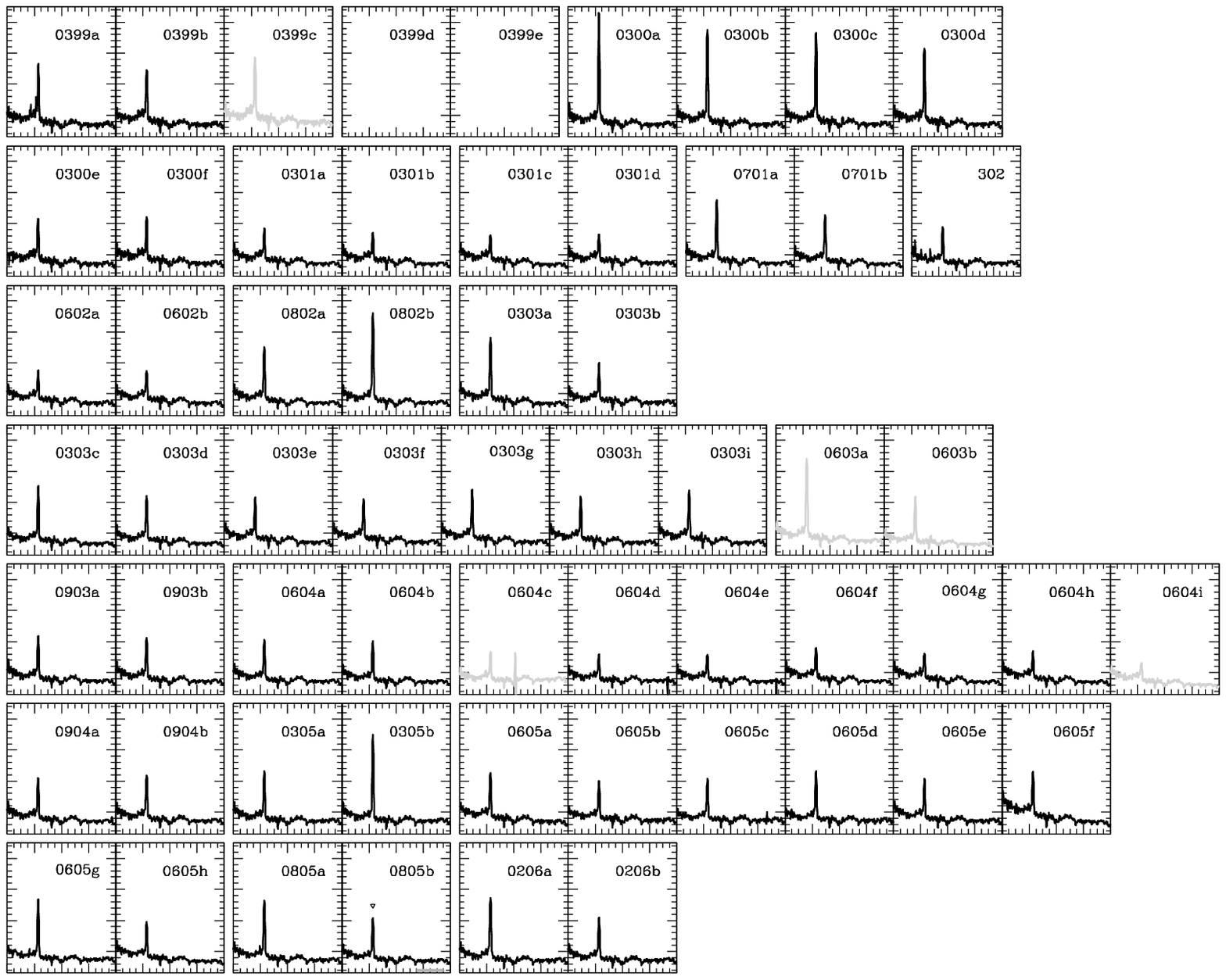,width=0.82\textwidth,angle=0} \\
\end{tabular}
\hspace*{100mm} 
\begin{minipage}{40mm} 
\vspace{-420mm} 
\caption{\footnotesize \ca\ region for each observation of Table~\ref{tab:obs}. 
In each window the horizontal axis goes from 3890 to 4012~\AA , and the vertical 
one from -0.05 to 1.2~$10^{-12}$~erg~cm$^{-2}$~s$^{-1}$~\AA$^{-1}$. Some 
spectra (0603a and b; 0604a, b, c, and i; and 0605g and h) are 
deficient and are not considered in the subsequent analysis. The first three 
spectra were acquired with a different instrumental configuration and they do 
not include this region.} 
\label{fig:CaProx}
\end{minipage}
\hspace*{135mm}
\begin{minipage}{40mm} 
\vspace{-190mm} 
\caption{\footnotesize \hal\ region for each observation of Table~\ref{tab:obs}. 
In each window the horizontal axis goes from 6540 to 6620~\AA , and the vertical 
one from  0.15 to 2.25~$10^{-12}$~erg~cm$^{-2}$~s$^{-1}$~\AA$^{-1}$. Some  
spectra (0399c, 0603a and b, and 0604c and i) are deficient and are not 
considered in the subsequent analysis. The fourth and fifth spectra were 
aqcuired with a different instrumental configuration and they do not include 
this region.} 
\label{fig:HaProx} 
\end{minipage} 
\end{figure*}

Since Proxima is a flare star, it is expected to have frequent flares. 
Therefore, before exploring the existence of an activity cycle in Proxima 
Centauri, we have to determine whether any flare occurred during our 
observations, since flare activity can mask the variations of activity due to 
the cycle, and it has to be filtered out before the analysis. 
In Figs.~\ref{fig:CaProx} and \ref{fig:HaProx} we show the line profiles 
of the \ca\ H and K and \hal\ lines. Visual inspection of the temporal series of 
observations reveals the possible occurrence of several flares. The first one 
corresponds to the four observations 0300a to 0300d: the apparent flare goes 
from maximum to minimum, with the two central observations with similar levels 
of activity. Also, in the spectrum 0802b, all the lines are much stronger than 
in the previous one 0802a. Similar variations can be appreciated between 
observations 0303a and 0303b, 0305b and 0305a, and 0206a and 0206b, although in 
these cases the lines are not so intense. 
Using these considerations, we separated the flaring spectra from the 
``normal'' ones. We excluded these flaring spectra --~which are 
indicated in Table~\ref{tab:obs} with a `$\star$'~-- from the rest of the  
analysis.

\section{Behavior of \hal\ with time}\label{sec:halfa}

\begin{table}[h!]
\caption{Nightly averaged $A$ index. Column 1: the label used 
in Fig.~\ref{fig:serie}; Col.~2: the xJD at the 
beginning of the observations; Col.~3: the quantity of spectra averaged;  
Col.~4: the total exposure time (in minutes); Col.~5: the average index.} 
\footnotesize \label{tab:A}\begin{center}
\begin{tabular}{lrcrr}\hline \hline
Label & \multicolumn{1}{c}{xJD} & $N$ & \multicolumn{1}{c}{\emph{t}} & \multicolumn{1}{c}{$A$} \\ \hline
0399    &    241.85 & 2 &  50 & 2.89 \\
0300    &    627.82 & 2 &  90 & 2.50 \\
0301\_1 &    972.78 & 2 & 120 & 2.14 \\
0301\_2 &    974.70 & 2 & 180 & 2.00 \\
0701    & 1\,096.46 & 2 & 180 & 2.84 \\
0302    & 1\,364.69 & 1 &  45 & 2.16 \\
0602    & 1\,451.66 & 2 &  90 & 2.13 \\
0802    & 1\,519.50 & 1 &  45 & 2.84 \\
0303\_1 & 1\,715.75 & 1 &  45 & 2.28 \\
0303\_2 & 1\,716.60 & 7 & 420 & 2.63 \\
0903    & 1\,895.51 & 2 &  66 & 2.46 \\
0604\_1 & 2\,160.59 & 2 &  90 & 2.35 \\
0604\_2 & 2\,161.51 & 5 & 300 & 2.00 \\
0904    & 2\,274.48 & 2 &  60 & 2.48 \\
0305    & 2\,448.78 & 1 &  45 & 2.61 \\
0605\_1 & 2\,523.55 & 6 & 360 & 2.46 \\
0605\_2 & 2\,543.60 & 2 & 180 & 2.54 \\
0805    & 2\,600.51 & 2 & 160 & 2.66 \\
0206    & 2\,780.86 & 1 &  80 & 2.42 \\ \hline
\end{tabular}

\end{center}
\end{table} 

\begin{figure}[b!]
\centering
\includegraphics[width=\columnwidth]{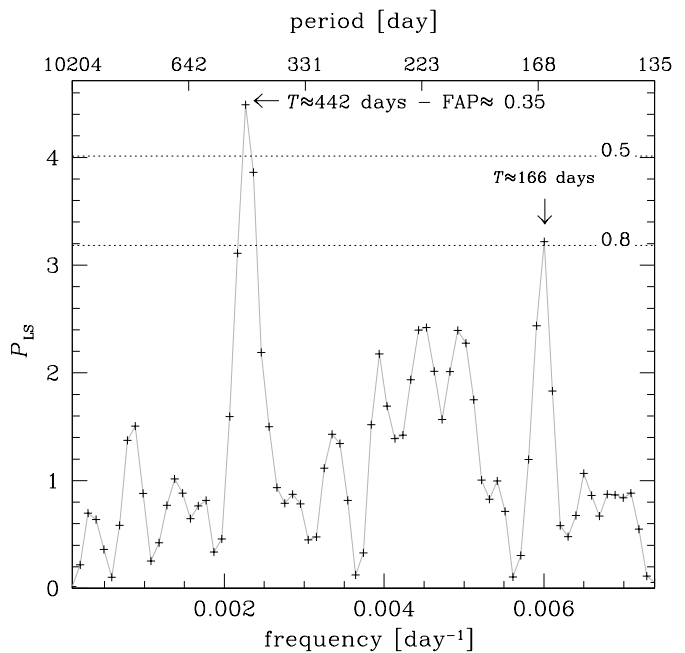}
\caption{Lomb-Scargle periodogram of the data of Table~\ref{tab:A}. The 
False Alarm Probability levels of 50 and 80\% are shown.} 
\label{fig:perio} 
\end{figure}

For the non-flaring spectra, we calculated a nightly average of the $A$ index, 
shown in Table~\ref{tab:A}, which we examine in this section to determine the 
possible presence of an activity cycle. 
Using this data, we first calculated the Lomb-Scargle periodogram 
\citep{1982ApJ...263..835S,1986ApJ...302..757H} using the algorithm included in 
\citet{1992nrca.book.....P}. This periodogram is a method to estimate the 
power spectrum (in the frequency domain), when observation times are 
unevenly spaced, and normalizes the spectrum in such a way that it is 
possible to measure the significance of the peaks. The amplitude of the 
periodogram at each frequency point is identical to the equation that would be 
obtained  estimating the harmonic content of a data set, at a given frequency, 
by linear least-squares fitting to an harmonic function of time 
\citep{1992nrca.book.....P}. We show this periodogram in Fig.~\ref{fig:perio}. 

It can be seen in the figure that there is a distinct peak at 442 
days, with a significance of 65\%. In other words, a period of 442 days is 
detected, with a \emph{False Alarm Probability} (FAP) of 35\%. It is worth 
noting that this FAP is very good, since the activity cycles are neither 
harmonic nor exactly periodic. 
Other less significant peaks are present, the most important one corresponding 
to 166 days with a FAP of 80\%. This peak could correspond to a subharmonic of a 
year, which can appear due to the fact that our observations are carried out 
with a periodicity of approximately one year.

We also applied several techniques in the time domain to check the existence of 
this period. Basically, all of them start with the construction of the light 
curve (that is, the observed data as a function of phase for different trial 
periods). Then, this light curve is partitioned in several bins, whose quantity 
depends on the amount of data available. In the three cases, we chose a 
partition of two bins, due to the small number of data points.

The first of these techniques is the one outlined in 
\citet{1971Ap&SS..13..154J}, which assumes that the real period is the one that 
minimize the dispersion, measured from the average, of the combined bins of the 
light curve. The second one \citep{1980PASP...92..700M} is an improvement of the 
first, which consists in measuring the dispersion relative to the least-squares 
straight line for each bin. Finally, \citet{1995ApJ...449..231C} instead of 
minimizing the dispersion, calculated the so-called Shannon entropy, which is 
a measure of the order of the data. Again, the real period will minimize this 
quantity. 

\begin{figure}[t!]
\centering
\includegraphics[width=\columnwidth]{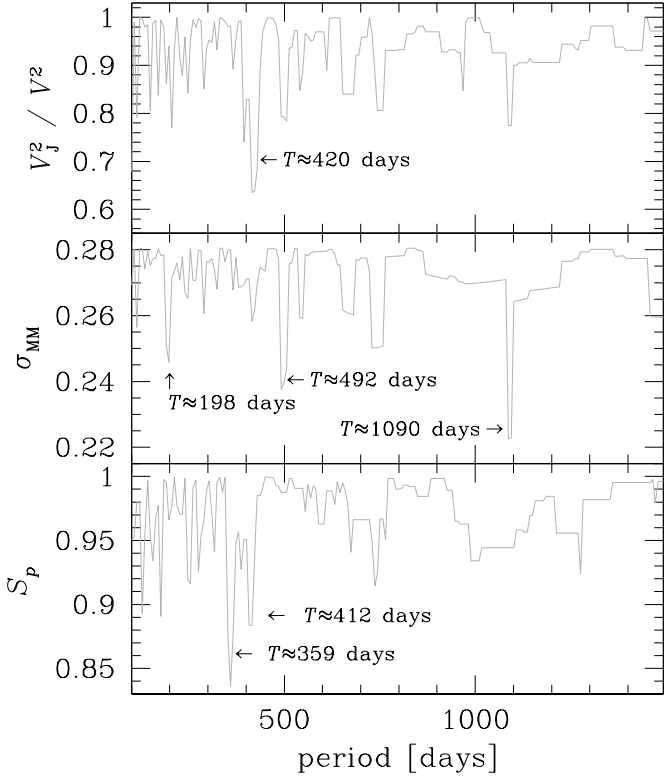}
\caption{The results of the three different methods employed that involve 
light-curve dispersion measurements: in the higher panel, the relative 
\citet{1971Ap&SS..13..154J} dispersion; in the middle panel, the 
\citet{1980PASP...92..700M} dispersion; and in the lower panel, the Shannon 
entropy \citep{1995ApJ...449..231C}, for the data of 
Table~\ref{tab:A}.} 
\label{fig:entropia} 
\end{figure}

In Fig.~\ref{fig:entropia} we show the dispersion of the light curve --~or the 
Shannon entropy~-- as a function of the trial period for each method. 
In all of them we find periods compatible within 10\% with the one of 
$\sim$442~days (420, 492, and 412 days). Also present are possible 
harmonics of a year (198, 359, and 1090~days). One should keep in mind that 
solar activity is not exactly periodic, since each cycle has a different length 
and intensity. Therefore, we believe that the agreement found between different 
methods is quite remarkable. 

\begin{figure}[b!]
\centering
\begin{tabular}{c}
\includegraphics[width=\columnwidth]{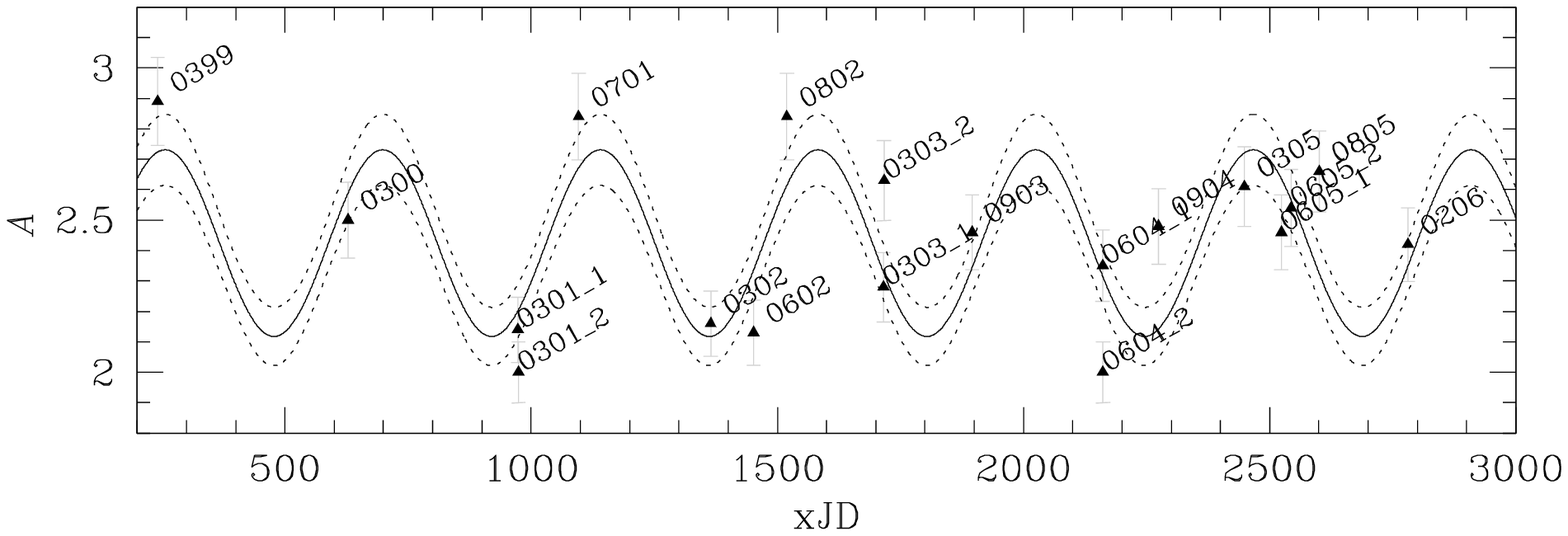} \\
\includegraphics[width=\columnwidth]{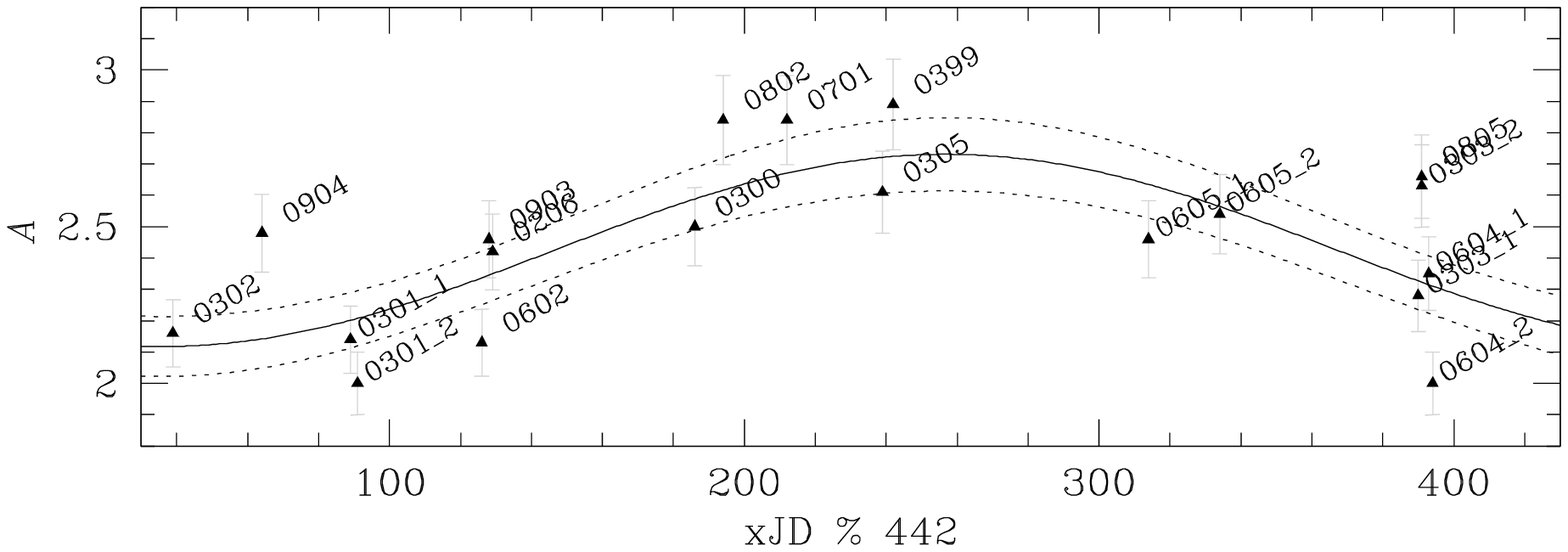}
\end{tabular}
\begin{minipage}[l]{\columnwidth}
\vspace{-120mm} \hspace{0mm} (\emph{a})
\end{minipage} 
\begin{minipage}[l]{\columnwidth}
\vspace{-65mm} \hspace{0mm} (\emph{b})
\end{minipage} 
\caption{\emph{a}) $A$ index \emph{vs.} time, assuming errors of 5\%  
for each point. The full curve is the least-squares fit to the data of 
Table~\ref{tab:A} after establishing a harmonic function of period 
442~days. The dotted curves represent the $\pm2\sigma$ deviations. \emph{b}) 
Light curve of the same data.}  
\label{fig:serie} 
\end{figure}

We also fit the data, by least-squares, with a harmonic function with a 
period of 442~days, and we obtained the curve shown in 
Fig.~\ref{fig:serie}\emph{a}. Also the fit is remarkably good here, again 
considering that the cycle is not exactly harmonic. Finally, in 
Fig.~\ref{fig:serie}\emph{b} we draw the associated light curve.                
                                                                 
As measured from minimum to maximum, the amplitude of the period is 
around 25\% in $A$. If we translate this amplitude to the $S$ index, using the 
slope of Fig.~\ref{fig:AvsS}, we find a variation of about 130\% in a cycle, to 
be compared with a solar variation for $S$ of only $\sim$30\% 
\citep{1995ApJ...438..269B}. This much larger variability is consistent with the 
fact that Proxima Centauri is an extremely active star. Since there are no 
systematic studies of activity variations for M stars so far 
\citep[e.g., see][]{1995ApJ...438..269B}, we cannot compare with any other 
similar star. 

Finally, we checked the accuracy of the result of Fig.~\ref{fig:perio} 
using the sunspot numbers taken from the National Geophysical Data Center. 
To do so, we took a sample of the solar data with the same phase intervals that 
we have in our data for Proxima, we added Gaussian noise with errors of 
10\% at each point, and we computed the Lomb-Scargle periodogram. We repeated 
this procedure 1000 times with random starting dates.

\begin{figure}[t!]\centering
\includegraphics[width=\columnwidth]{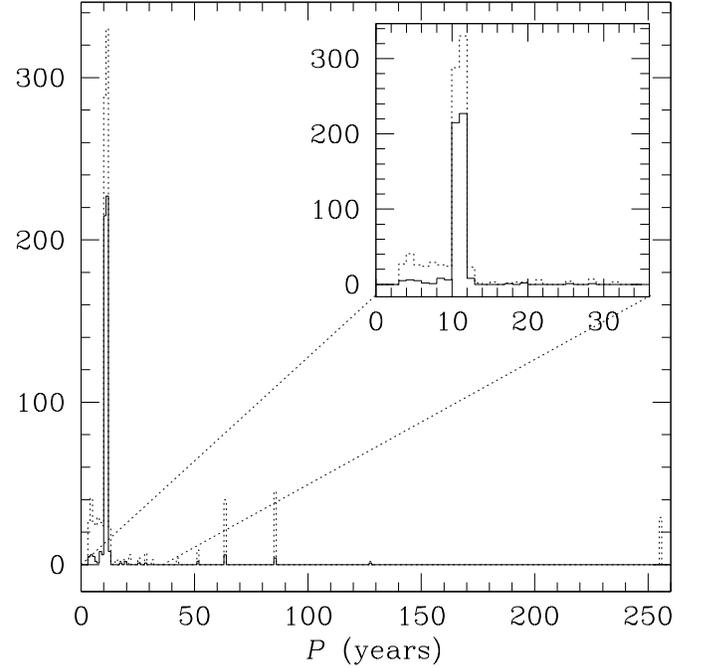}
\caption{Histogram of the detected periods (the periods for which we found the 
maxima $P_{\mathrm{LS}}$) for a sample of 1000 different sets taken from the 
Sun. The full line shows only the periods with FAP~$\leq 0.35$ and the dotted 
line shows all the cases.} 
\label{fig:histosol} 
\end{figure}

For each periodogram, we considered the period with 
maximum significance (or minimum FAP) as the detected period . In 
Fig.~\ref{fig:histosol} we plot the histogram of these 
detected periods. The full line shows only the periods with FAP~$\leq 0.35$. We 
see that the correct period (10 to 12 years) was detected in 62\% of the 
periodograms and in 44\% of the cases with a FAP~$\leq 0.35$. Only in 6\% of the 
cases was a ``false'' period detected with a FAP~$\leq 0.35$. 

Therefore, if this star has a cyclic behavior similar to the solar one, the 
probability of detecting it in our observations is $P\! \sim\!  60\%$ and the 
probability of detecting it with a FAP~$\leq 0.35$ is $P\! \sim\!$  
45\%. On the other hand, if a period is detected with a FAP~$\leq 
0.35$, its value is correct with a probability of $P\! \sim\! 90\%$.

\section{Discussion} \label{sec:conclusiones}

In this paper we found strong evidence of a cyclic activity in the dMe star 
Proxima Centaury, with a period of $\sim$442 days $\pm 10\%$. Similar values for 
the period were found from the Lomb-Scargle periodogram and using three 
different techniques in the time domain. We were also able to determine that the 
activity variations outside of flares amount to 130\% in $S\!$, four times 
larger than for the Sun. 

As we mentioned before, the possible existence of an activity cycle in this star 
is mentioned in three previous papers. First, \citet{1990A&A...232..387H} found 
some evidence \emph{suggesting} a stellar cycle, although they did not estimate 
its period. \citet{1996AAS...188.7105G} found some manifestation that in 
1995 Proxima was near a minimum of activity, again without an estimate of its 
period. In a more detailed study, \citet{1998AJ....116..429B} found some 
hints of an activity cycle of about 1100~days, compatible with the minimum 
activity around 1995, although they had some gaps in temporal coverage. The 
middle point of what they consider minimum activity corresponds to JD 
2\,449\,750. If we extrapolate our period back in time, taking into account that 
in JD 2\,451\,500 we have minimum activity, we find that in JD 2\,449\,732 we 
also have minimum activity, very close to what 
\citet{1998AJ....116..429B} found. 
However, the data used in these two last works were also used to 
study the rotational period of this star, giving very different results: 
\citet{1996AAS...188.7105G} found \prot=31.5$\pm$1.5~days and 
\citet{1998AJ....116..429B} determined \prot=83.5~days. On the other 
hand, \citet{1999A&A...344L...5K} did not find any evidence for the rotation 
periods found in these two works, based on very precise radial velocities of 
Proxima Centauri, which they analyze using both the Scargle periodogram and a 
sine-fitting routine, which minimize $\chi^2$ taking into account data errors. 
Therefore, we believe that our value of $\sim$442 days for the cycle is more 
reliable, in particular because all the methods we employed give similar values 
for the period, and because we estimated, using solar data, that if a cycle is 
present, the period we found is the correct one with a probability $P\sim 90\%$. 

Concerning the flare frequency, \citet{1981MNRAS.195.1029W} predicts a flare 
of intensity greater than $5\,10^{27}$ in the $U$ band once every 0.8~hr. For 
the most intense flares, the predicted frequency falls to one every 31~hr. In 
our case, we observed Proxima for $\sim$54~hr, detecting two very strong flares 
(0300a to d and 0802b) and three weaker ones (0303a, 0305b, and 0206a) in this 
period. Therefore, we found a frequency of one flare every 10~hr.

It is generally accepted that magnetic activity results from the generation of a 
large-scale toroidal field by the action of differential rotation on a poloidal 
field, at the interface between the convective envelope and the radiative core 
\citep{1975ApJ...198..205P,1992A&A...265..106S}. This is the $\alpha\Omega$ 
dynamo, which predicts a strong correlation between activity and rotation, which 
is in fact observed for stars from F to early M. This theory also explains the 
fact that young rapid rotators --~which have larger differential rotation~-- 
have higher activity levels than old slower rotators, with less differential 
rotation and therefore a less important dynamo effect. 

On the other hand, stars ranging from M3 to M9 are thought to be fully 
convective, and therefore they do not support an $\alpha\Omega$ dynamo. It 
has been argued \citep{2003ApJ...586..464B} that fully convective stars 
do not possess large-scale dynamos, and that their rotational morphology are 
dominated instead by small-scale turbulent fields. Nevertheless, there is plenty 
of observational evidence that slow late-type rotators like dMe stars have 
strong magnetic fields, with filling factors larger than for earlier stars and 
enhanced activity 
\citep{1980ApJ...239L..27M, 1989PhDT.........1H, 1996A&A...310..245M}. 

Recently,  \citet{2006A&A...446.1027C} showed that these cool objects can 
support large-scale magnetic fields by a pure $\alpha^2$ dynamo process. In this 
dynamo, helicity is generated by the action of the Coriolis force on the 
convective motions in a rotating, stratified fluid, and produces an  
amplification of the mean magnetic field.  Moreover, these fields can produce 
high levels of activity like the ones observed in M stars. This  
$\alpha^2$ dynamo does not predict a cyclic activity. However, our observations 
suggest that this cool star does have a clear period. 

The activity cycles measured in earlier stars (F to K) are all longer than 
2.5~years \citep{1995ApJ...438..269B}, to be compared to the one we found for 
Proxima Centaury, of only 1.2~years. On the other hand, the variation of the 
activity levels is much larger for this star than for earlier-type ones.
As we explained, the dynamo models show that the generation of activity 
in M stars is a very different phenomenon than in earlier stars, i.e., stars 
with a radiative core and an outer convection zone. It is desirable to confirm 
the existence of a cycle in Prox Cen and its period, and to explore other M 
stars near the limit where stars become fully convective, to constrain 
the dynamo at work in these stars. 


\begin{acknowledgements}
The CCD and data acquisition system at CASLEO has been partly financed by R. M. 
Rich through U.S. NSF grant AST-90-15827.  
\end{acknowledgements}

\bibliographystyle{aa}
\bibliography{prox}



\end{document}